\documentclass[aps,showpacs,groupedaddress,prl,twocolumn,nobalancelastpage,preprintnumbers]{revtex4}

\usepackage{graphics}      
\usepackage[pdftex]{graphicx,color}
\usepackage{natbib}
\usepackage{amsmath}
\usepackage{amssymb}
\usepackage{braket}
\usepackage{bbm}
\usepackage{units} 

\newcommand{\im}{i} 
\newcommand{\tr}{\textrm{Tr}}
\newcommand{\myO}{\mathcal{O}}
\newcommand{\Coh}{\mathcal{C}}
\newcommand{\Ent}{\mathcal{E}}
\newcommand{\myS}{\mathcal{S}}
\newcommand{\myD}{\mathcal{D}}

\begin{document}

\preprint{\textsf{published in Phys.~Rev.~Lett.~{111}, 030405 (2013)}}

\title{
Optimal coherent control to counteract dissipation
}

\author{Simeon Sauer}
\author{Clemens Gneiting\footnote{clemens.gneiting@physik.uni-freiburg.de}}
\author{Andreas Buchleitner}
\affiliation{Physikalisches Institut, Albert-Ludwigs-Universit\"at Freiburg, Hermann-Herder-Stra{\ss}e~3, D-79104 Freiburg, Germany}
\date{\today}

\begin{abstract}
We study to what extent the detrimental impact of dissipation on quantum properties can be compensated by suitable coherent dynamics. To this end, we develop a general method to determine the control Hamiltonian that optimally counteracts a given dissipation mechanism, in order to sustain the desired property, and apply it to two exemplary target properties: the coherence of a decaying two-level system, and the entanglement of two qubits in the presence of local dissipation.
\end{abstract}

\maketitle

Genuine quantum features such as entanglement or coherence are resources as precious as fragile, and their uncovering usually requires strong efforts in isolating and controlling quantum systems. Without thorough measures, decoherence efficiently shields the quantum world from our access and hides it behind its classical guise. While there has been unprecedented progress in the quantum control of various model systems, e.g. ions \cite{Haffner:2008}, quantum dots \cite{Hanson:2007}, or cold atoms \cite{Bloch:2008}, it is not possible to completely decouple these systems from their environment and thus to fully suppress the detrimental effect of decoherence. Standard optimal control techniques therefore focus on accessing quantum features in the transient regime, and the exploration and exploitation of quantum properties is consequently confined to a finite, generically short time window.

There are however ways to keep the window to the quantum world enduringly open, e.g. by encoding quantum features in topological properties of a system \cite{Nayak:2008}, or by engineering a dominant environment that drives the system into a highly nonclassical stationary state \cite{Diehl:2008,Verstraete:2009}. While these approaches in principle permit one to prepare arbitrary nonclassical quantum states, they require in general an exceedingly large overhead of resources.

In this letter we therefore ask to what extent already standard Hamiltonian control can enduringly counteract the detrimental effect of decoherence. Explicitly, we seek Hamiltonians that optimally uphold, on \emph{asymptotic} time scales, a given control objective (e.g., coherence, entanglement, or fidelity w.r.t. a target state) in the presence of dissipation. Such an asymptotic time behavior can be meaningfully formulated for static and, more generally, periodically time-dependent Hamiltonians. In the latter case, the asymptotic dynamics are periodic cycles in state space, which reduce to stationary states in the static case.

For our goal to single out the optimal among all conceivable control Hamiltonians, it is not advisable to directly scan the space of Hamiltonians, as the latter cannot efficiently be parametrized. We therefore approach the problem from a different perspective and determine the optimal stationary state or asymptotic cycle directly, i.e. independently of the Hamiltonian. The crucial insight behind is that physically admissible trajectories in state space are strongly constrained by the dissipative part of the dynamics. It thus turns out that one can characterize all possible stationary states or asymptotic cycles from the dissipative dynamics alone.

While our approach can be applied to the optimization of arbitrary control objectives, we demonstrate its viability with two physically relevant examples: the coherence between ground and excited state of a decaying two-level system, and entanglement of two qubits in the presence of local dissipation.

\paragraph*{Static control Hamiltonians.}

We consider an open quantum system evolving under a Lindblad master equation \cite{Breuer:2007},
\begin{align}
\dot \rho(t) = \im [\rho(t),H(t)] + \myD(\rho(t)), \quad (\hbar=1) \label{eq:Lindbladeq}
\end{align}
with a dissipator $\myD(\rho) = \sum_{k} \gamma_k [ L_k \rho L_k^\dagger - \frac{1}{2}\{ L_k^\dagger L_k, \rho\}_+ ]$ 
composed of Lindblad operators $L_k$ and rates $\gamma_k$. For an arbitrary but fixed dissipator $\myD(\rho)$, our goal is to optimize the stationary state $\rho_\textrm{ss}$ (for static $H$) or asymptotic cycle $\rho_\textrm{ac}(t)$ (for periodic $H(t)$) of \eqref{eq:Lindbladeq} w.r.t. an arbitrary objective function $\myO(\rho)$. We emphasize that this definition of optimality differs from quantum control scenarios that aim at rapidly preparing a given target state on time scales when decoherence is negligible. The target states of such time-optimal protocols are reached quickly, but persist only on transient time scales, whereas the optimal cycles in our approach may take a long time to emerge, but then persist for arbitrarily long times.

It is instructive to investigate static control Hamiltonians first.
Direct optimization over all conceivable Hamiltonians $H$ involves the stationarity condition,
\begin{equation}
0 =\im [\rho_\textrm{ss},H] + \myD(\rho_\textrm{ss}) \label{eq:stationarity},
\end{equation}
in order to infer the stationary state $\rho_\textrm{ss}$ for a given $H$. Its inversion typically requires numerical means and must be repeated for each sample Hamiltonian, rendering this approach impractical already in low-dimensional systems. Therefore, we develop a different strategy here. Instead of starting from Hamiltonians, we base the optimization on the set of \emph{stabilizable states} $\myS$ \cite{Recht:}:
\begin{equation}\label{eq:DefS}
	\myS = \{ \rho : \ \exists H \ \textrm{s.t.} \  0= \im [\rho,H] + \myD(\rho) \}.
\end{equation}
It comprises all quantum states that become stationary under a suitable Hamiltonian. As shown below, this set can be characterized \emph{independently} of the Hamiltonian. Optimization of an objective function $\myO(\rho)$ can then be done in $\myS$ directly.

To derive this Hamiltonian-independent characterization of $\myS$, we exploit that the coherent dynamics induced by $H$ and the dissipative dynamics induced by $\myD(\rho)$ must compensate each other for a stationary state. Since the coherent part of the master equation \eqref{eq:Lindbladeq}, $\im [\rho,H]$, necessarily leaves the spectrum of $\rho$ invariant, this must also hold for the dissipator $\myD(\rho)$ at a stationary state. In other words, $\myD(\rho)$ does not modify the purity $p=\tr[\rho^2]$ or any higher moment $\tr[\rho^n]$ with $n>2$. This implies
$\left.\partial_t \tr[\rho^n] \right|_{H=0} \overset{\eqref{eq:Lindbladeq}}{=} n \tr[\rho^{n-1}\myD(\rho)] = 0$,
if $\rho$ is in $\myS$.
Since the moments of a $d$-dimensional quantum state are independent only up to $n=d$, this leads to $d-1$ necessary conditions for $\rho$ to be stabilizable:
\begin{equation}\label{eq:crit}
\rho \in \myS \quad \Rightarrow \quad \forall \, n \in \{2,\dots,d\}: \ \tr[\rho^{n-1}\myD(\rho)] =0.
\end{equation}
Denoting by $\myS_n$ the set of states that fulfill \eqref{eq:crit} for fixed $n$, we have $\myS\subset \bigcap_{n}\myS_n$.
For states with non-degenerate eigenvalues, criterion \eqref{eq:crit} is also sufficient \cite{Suppl:1}, and hence $\myS  =  \bigcap_{n}\myS_n$.
Note that this hierarchical characterization of $\myS$ does not require reference to the stabilizing Hamiltonian $H$, in contrast to definition \eqref{eq:DefS}. Given a stabilizable state $\rho\in\myS$, however, it is straightforward to derive the corresponding $H$ from \eqref{eq:stationarity}, based on the spectral decomposition $\rho=\sum_\alpha \lambda_\alpha \ket{\alpha}\bra{\alpha}$:
\begin{equation}\label{eq:HgivenRho}
H=\sum_{\alpha,\beta: \lambda_\alpha \ne \lambda_\beta } \frac{i \braket{ \alpha | \myD(\rho) | \beta} }{\lambda_\alpha - \lambda_\beta} \ket{\alpha}\bra{\beta}.
\end{equation}

\begin{figure}[tb]
\begin{tabular}{ l l }
	(a) ${\gamma_+}/{\gamma_-} = {\gamma_d}/{\gamma_-} =0$ & (b) ${\gamma_+}/{\gamma_-} = {\gamma_d}/{\gamma_-} ={1}/{4}$ \\
	\includegraphics[width=.4\linewidth]{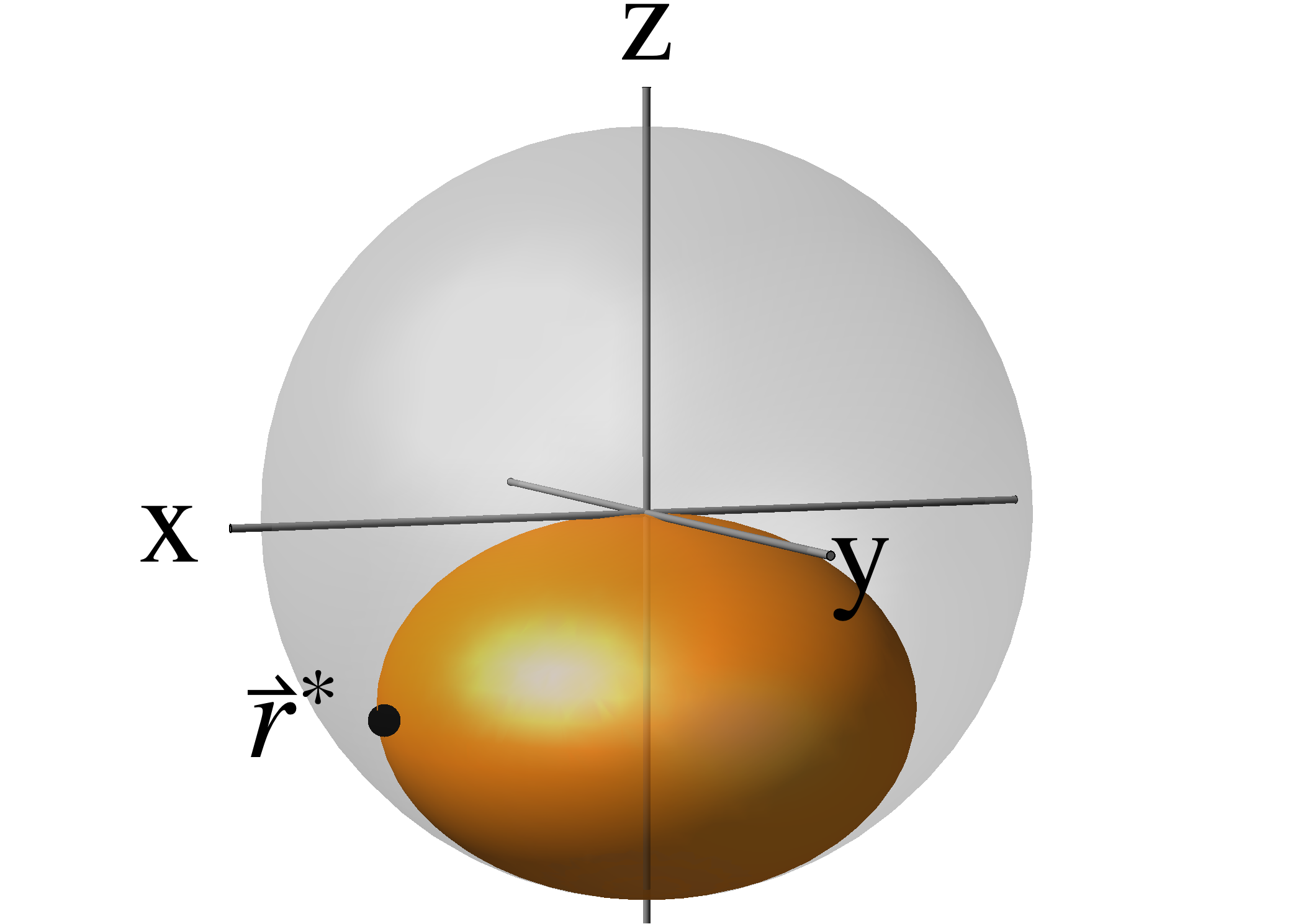} &
	\includegraphics[width=.4\linewidth]{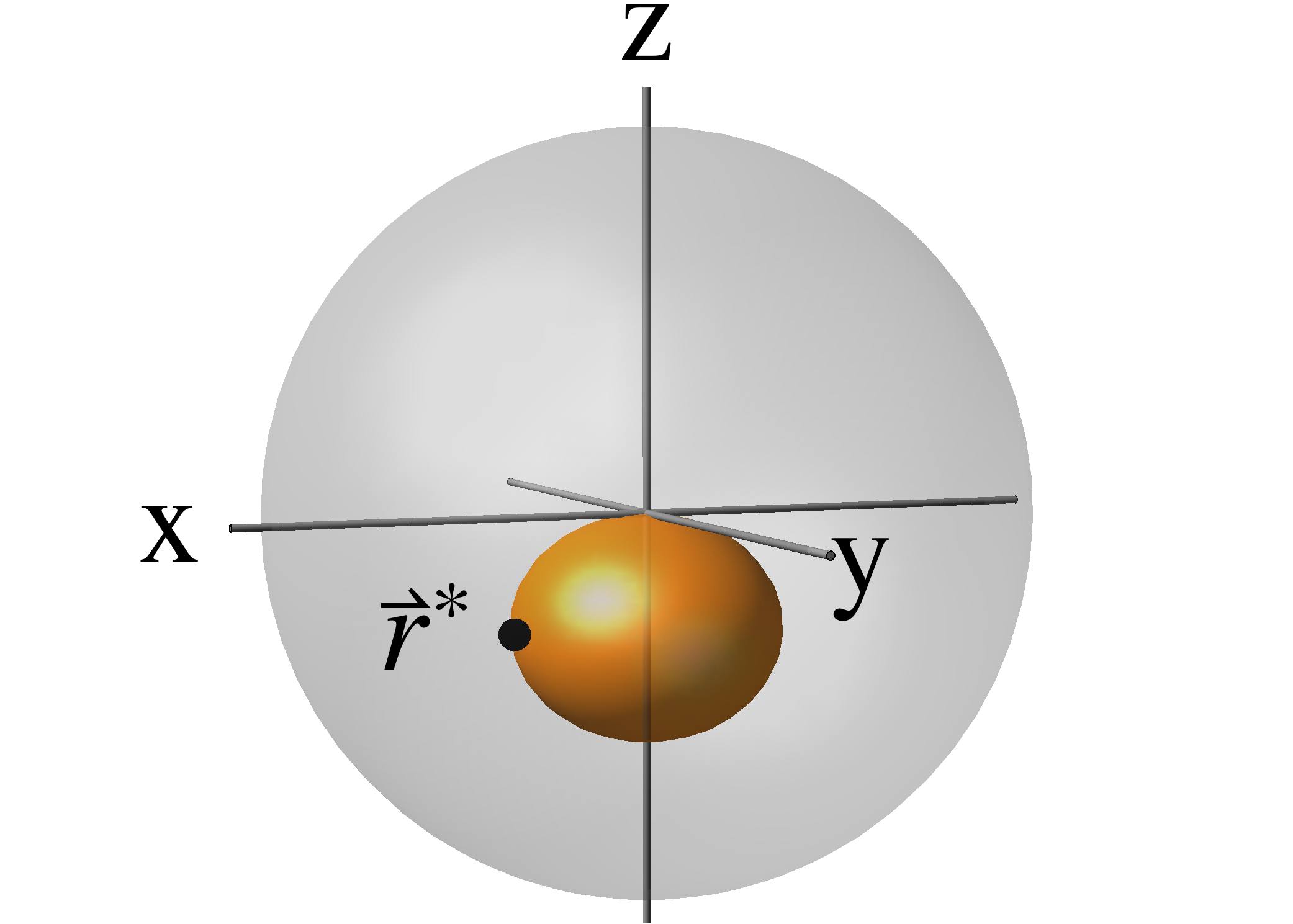}
\end{tabular}
\caption{Bloch representation of the set $\myS$ of stabilizable states (surface of the orange (dark shaded) spheroid) for a single qubit, subject to spontaneous decay, excitation, and dephasing. The respective rates $\gamma_-$, $\gamma_+$, and $\gamma_d$ determine the shape of $\myS$. (a) For spontaneous decay only, the optimal stationary state $\vec{r}_{*}$ w.r.t.~to the coherence $\Coh$ (i.e., the distance to the $z$-axis) reaches $\Coh={1}/{\sqrt{2}}$. (b) With finite $\gamma_+$ and $\gamma_d$, the ellipsoid is contracted, reducing the optimal value of $\Coh$.}\label{fig:ellipsoid}
\end{figure}

To demonstrate the viability of our method, we first discuss the case of a single qubit. There, one finds an intuitive geometric representation of $\myS$. Since $d=2$, \eqref{eq:crit} imposes merely a single constraint ($n=2$). In terms of the Bloch vector, $\vec{r}=\tr[\rho \vec{\sigma}]$, this constraint defines a quadric hypersurface in the Bloch ball:
\begin{equation}\label{eq:quadric}
\vec{r}\in\myS \quad \Rightarrow \quad \vec{r} \cdot \left(D \vec{r} + \vec{d} \right) = 0.
\end{equation}
Here, the $3\times 3$ matrix $(D)_{ij}=\tr[\sigma_i \myD(\sigma_j)]$ and the vector $(\vec{d})_i=\tr[\sigma_i \myD(\mathbbm{1})]$ characterize the dissipator in Bloch notation. According to \eqref{eq:quadric}, a state $\vec{r}$ is stabilizable, if (and only if \footnote{The ``only-if'' holds for any state apart from the totally mixed state $\vec{r}=\vec{0}$, which has degenerate eigenvalues, so that criterion \eqref{eq:quadric} is not sufficient for stabilizability. However, $\vec{0}$ lies in the closure of $\myS$, as apparent from Fig.~\ref{fig:ellipsoid}. We strongly conjecture that $\bigcap_{n}\myS_n$ is \emph{always} identical to the closure of ${\myS}$, since states with non-degenerate spectra are dense in the state space, making it plausible that they are dense in $\myS$ as well.}) the dissipative flux $D \vec{r} + \vec{d}$ is orthogonal to $\vec{r}$, i.e., if it has no radial component.

Specifically, we consider a qubit exposed to the three most common incoherent processes: decay of the excited state at rate $\gamma_-$, absorption from the ground state at rate $\gamma_+$, and dephasing between ground and excited state at rate $\gamma_d$. Experimental realizations of this scenario include both atomic and solid-state two-level systems, such as trapped ions \cite{Leibfried:2003}), superconducting qubits \cite{Chiorescu:2003}, or color centers in diamond \cite{Buckley:2010}. Specifically, in the latter case, the incoherent processes are triggered by a nuclear spin bath, resulting in typical incoherent rates in the kHz regime \cite{Ryan:2010}.

With these particular incoherent processes, $\myS$ is the surface of a spheroid \cite{Recht:}, with the polar axis pointing in $z$-direction, cf.~Fig.~\ref{fig:ellipsoid}. The polar and equatorial diameter depend on the incoherent rates. The optimal stationary state w.r.t. an arbitrary objective $\myO(\vec{r})$ is now conveniently determined by maximizing over the surface of this spheroid. E.g., one may consider the coherence $\Coh=2|\braket{0| \rho |1}|$ between ground and excited state.
In Bloch notation, this objective corresponds to the distance to the $z$-axis, see Fig.~\ref{fig:ellipsoid}. Hence, the optimal coherence equals the equatorial semi-axis of the spheroid, yielding $\Gamma_-/2 \Omega$, with $\Omega=\sqrt{\Gamma_+({\Gamma_+}/{2}+\gamma_d)}$ and $\Gamma_\pm=\gamma_-\pm \gamma_+$.
According to \eqref{eq:HgivenRho}, the corresponding Hamiltonian reads $H^*=-(\Omega/2)\sigma_y$. It can be realized, e.g., by resonantly driving the qubit with a Rabi frequency $\Omega$ \cite{Cohen-Tannoudji:1998}.

As a higher-dimensional example, we consider two qubits ($d=4$). The stabilizable states $\myS$ then lie in the intersection of three hypersurfaces $\myS_2$, $\myS_3$, and $\myS_4$.
Similarly to a single qubit, one can represent $\rho$ by a 15-dimensional Bloch vector $\vec{r}$ \cite{Bengtsson:2006,Byrd:2003}.
The lowest order constraint ($n=2$) can then be cast into the same form as \eqref{eq:quadric}, defining again a quadric surface.
The higher order constraints for $n=3,4$, however, lead to polynomial expressions of third and fourth degree in $\vec{r}$. Therefore, instead of determining the optimal state in $\myS$ directly, it is favorable to determine the optimal state $\vec{r}_{*}$ in $\myS_2$ first, and then to verify that $\vec{r}_{*}$ lies also in $\myS$. If so, it must be the optimum in $\myS$, since $\myS \subset \myS_2$. If not, the procedure provides an upper bound for the optimal value in $\myS$.

Relevant target properties for two qubits address, e.g., their entanglement. The most detrimental situation is then certainly encountered when the incoherent processes act locally on both qubits. Therefore, we exclusively consider qubits undergoing (individual) spontaneous decay at rate $\gamma_-$, as encountered in the experimental scenarios mentioned above for a single qubit.

As a specific entanglement objective, we study the fidelity $\mathcal{F}(\rho)=\braket{\Psi_+|\rho|\Psi_+}$ w.r.t. the maximally entangled Bell state $\ket{\Psi_+}=(\ket{01}+\ket{10})/\sqrt{2}$. It is relevant, e.g., in teleportation protocols \cite{Horodecki:1999}.
It optimization over $\myS_2$ can be carried out analytically, leading to the optimal stationary state
\begin{equation}\label{eq:rhoopti}
\rho^{*} = ({1}/{2}) \ket{00}\bra{00} +  ({1}/{2}) \ket{\Psi_+}\bra{\Psi_+},
\end{equation}
with $\mathcal{F}(\rho^*)={1}/{2}$. The Hamiltonian $H(\alpha,\beta) = \mathbbm{1} \otimes \left( \alpha \sigma_z +\beta \sigma_x \right) + \left(\alpha \sigma_z + \beta \sigma_x\right)  \otimes \mathbbm{1} - 2 \alpha ( \sigma_+ \otimes \sigma_-  + \sigma_- \otimes \sigma_+)$
stabilizes $\rho^{*}$ in the limit of $\beta / \gamma_- \rightarrow \infty$ and $\alpha /  \beta \rightarrow \infty$ \footnote{Thus, strictly speaking, $\rho^{*}$ is not in $\myS$, but in its closure, consistent with the fact that is has degenerate eigenvalues.}.
This Hamiltonian is readily realized in various experimental setups; in particular, the interaction term describes an excitation hopping mechanism, realizable with trapped ions \cite{Haffner:2008}, superconducting circuits \cite{Steffen:2006}, dipole-dipole interactions between excitons \cite{Forster:1948,Adolphs:2006}, color centers in diamond \cite{Neumann:2010}, and Rydberg atoms \cite{Gallagher:2008mz,Gurian:2012}.

Another relevant two-qubit objective is the entanglement measure concurrence $\Ent(\rho)$ \cite{Wootters:1998}. In contrast to the fidelity, it does not favor a specific state, but assigns full concurrence $\Ent=1$ to all maximally entangled states. Since $\Ent(\rho)$ is not linear in $\rho$, however, its optimization cannot be treated analytically. Numerical optimization in $\myS_2$ reveals that the optimal state coincides with the fidelity-optimal state $\rho^{*}$ of Eq.~\eqref{eq:rhoopti}, yielding $\Ent(\rho^{*})=1/2$. This is plausible, since a high Bell state fidelity typically implies strong entanglement. W.r.t. the concurrence, however, $\rho^*$ is not the unique optimum; e.g., \eqref{eq:rhoopti} with $\Psi_+$ replaced by $\Psi_-$ is a stabilizable state with concurrence $1/2$, as well.
 
These results clearly indicate the possibilities and limitations of coherent control of open systems: on the one hand we proved the impossibility of exceeding the fifty-fifty mixture \eqref{eq:rhoopti} of the deexcited state and a maximally entangled Bell state; on the other hand, $\Ent=1/2$ is still significantly higher than the average concurrence $\Ent = 0.18$ of typical stationary states, as we have verified by a statistical sampling of Hamiltonians.

\paragraph*{Periodic control Hamiltonians.}

So far, we developed a method to determine the static Hamiltonian $H$ that upholds the optimal amount of an objective $\myO$ in the stationary state.
We now extend these concepts to the envisaged, substantially more general case of periodic control Hamiltonians $H(t)=H(t+T)$.
Since this comprises static Hamiltonians as a special case, one expects this additional freedom in control to improve the optimization results.

Specifically, our aim is to determine periodic Hamiltonians $H(t)$ which optimize $\overline{\myO} = \frac{1}{T}\int_0^T \myO(\rho_\textrm{ac}(t)) \mathrm{d}t$, i.e., the time average of the objective $\myO$ in the asymptotic cycle $\rho_\textrm{ac}(t)$. To this end, we optimize $\overline{\myO}$ in the set $\mathcal{A}$ of \emph{stabilizable cycles}, comprising all periodic trajectories $\rho(t)=\rho(t+T)$ (with arbitrary period $T$) for which a periodic $H(t)$ exists such that $\rho(t)$ solves the master equation \eqref{eq:Lindbladeq}.
Criterion \eqref{eq:crit} is then generalized to
\begin{align}
\forall t \ \forall n: \tr[\rho^{n-1}(t)\myD(\rho(t))] = \frac{1}{n}\partial_t \tr[\rho(t)^n], \label{eq:crit2}
\end{align}
which holds for any $\{\rho(t)\} \in \mathcal{A}$. Eq.~\eqref{eq:crit2} reflects the fact that only the dissipative term $\myD(\rho)$ can change the spectrum of $\rho$ and thus its spectral moments $\tr[\rho^n]$.
As before, criterion \eqref{eq:crit2} is also sufficient for $\{\rho(t)\}\in\mathcal{A}$, if $\rho(t)$ has non-degenerate eigenvalues for all $t$. The Hamiltonian $H(t)$ stabilizing a given cycle in $\mathcal{A}$ is found analogously to \eqref{eq:HgivenRho}.

E.g., for the purity $p$ (i.e., $n=2$), \eqref{eq:crit2} implies that a consistent cycle must equally probe regions of the state space where the \emph{purity flux} $f(\rho)\equiv \tr[\rho\myD(\rho)]$ of the dissipator is positive and regions where it is negative.
The regions of positive and negative purity flux are separated by the hyperplane of vanishing flux, i.e., by the set $\myS_2$ introduced in the discussion of the static case. Hence, any cycle must intersect  $\myS_2$ an even number of times (at least twice). Moreover, no cycle can explore regions where the purity is larger than the maximal purity in $\mathcal{S}_2$ \cite{Suppl:2}.

It is again instructive to consider a single qubit first. A cycle $\{\rho(t):t\in[0,T)\}$ is then represented by a closed trajectory $\{\vec{r}_{t}\}$ in the Bloch ball, and stabilizable cycles are characterized by criterion \eqref{eq:crit2} with $n=2$, reading
\begin{equation}\label{eq:purefluxBloch}
\underbrace{ \vec{r}_{t} \cdot \left(D \vec{r}_{t} + \vec{d} \right)}_{\equiv f(\vec{r}_t)} = \underbrace{\frac{1}{2}\partial_t |\vec{r}_{t}|^2}_{ \equiv \dot p(\vec{r}_t)}.
\end{equation}
It reflects the fact that the time evolution of the purity $p=\tr[\rho^2]\equiv(|\vec{r}|^2+1)/2$ is exclusively governed by the radial part $f(\vec{r})$ of the dissipator.

\begin{figure}[tb]
\includegraphics[width=1\linewidth]{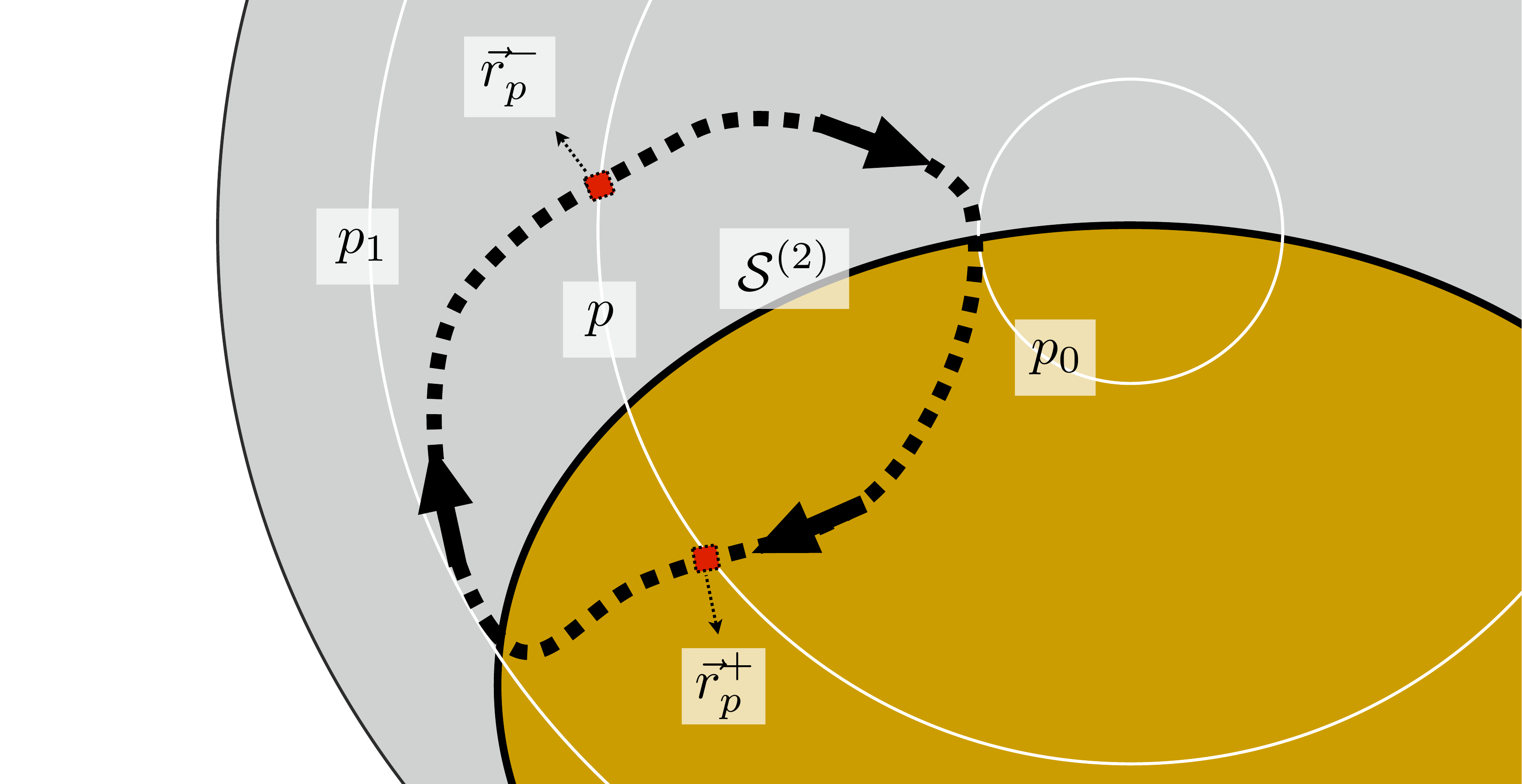}
\caption{Illustration of a general asymptotic cycle $\{\vec{r}_{t}\}$ (dotted loop), and a two-point cycle $\{ (\vec{r}^+_p,\vec{r}^-_p)\}$ (red boxes). White rings connect points of equal purity $p$. The region of positive purity flux is orange (dark gray). It is separated from the region of negative flux (light gray) by the hypersurface $\myS_2$ (solid black line) of states that can be stabilized by static Hamiltonians.}\label{fig:TwoPointCycle}
\end{figure}

While an optimization over all stabilizable cycles is certainly unfeasible, the problem can be reduced to the tractable class of \emph{two-point cycles} (TPC), based on the following argument: Any cycle undergoes subsequent stages of strictly monotonic purity gain and loss. Without loss of generality, we consider cycles that consists of two stages, intersecting $\myS_2$ twice; general cycles reduce to this case. To each point $\vec{r}_p^+$ in the purity-increasing stage ($f(\vec{r}_p^+)>0$), one can assign a point $\vec{r}_p^-$ of equal purity $p$ in the purity-decreasing stage ($f(\vec{r}_p^-)<0$). Hence, the cycle can by parametrized by $p$, see Fig.~\ref{fig:TwoPointCycle}.
Denoting by $p_0$ ($p_1$) the minimal (maximal) purity of the cycle and using \eqref{eq:purefluxBloch}, we find that the objective $\overline{\myO}$ for an arbitrary cycle is majorized by a TPC,
\begin{equation}\label{eq:estimation}
\overline{\myO} = \frac{ \int_{p_0}^{p_1} \left( \frac{\myO(\vec{r}_p^+)}{|f(\vec{r}_p^+)|} + \frac{\myO(\vec{r}_p^-)}{|f(\vec{r}_p^-)|} \right) \mathrm{d}p }{ \int_{p_0}^{p_1} \left( \frac{1}{|f(\vec{r}_p^+)|} + \frac{1}{|f(\vec{r}_p^-)|} \right) \mathrm{d}p } \le \max_{p} \overline{\myO}_\textrm{TPC}(\vec{r}_p^+,\vec{r}_p^-).
\end{equation}
Here, $\overline{\myO}_\textrm{TPC}$ defines the time-averaged objective of a TPC:
\begin{equation}\label{eq:OTPC}
\overline{\myO}_\textrm{TPC}(\vec{r}_p^+,\vec{r}_p^-) = \frac{\myO(\vec{r}_p^+)| f(\vec{r}_p^-)|+\myO(\vec{r}_p^-)| f(\vec{r}_p^+) | }{| f(\vec{r}_p^+)|+| f(\vec{r}_p^-) | }.
\end{equation}
In \eqref{eq:estimation}, we used the estimate $\int_{a}^{b} g(x) \mathrm{d} x / \int_{a}^{b} w(x) \mathrm{d} x \le \max_{x} g(x)/w(x)$, holding for any positive $w(x)$ and arbitrary $g(x)$.
The cycle that achieves $\overline{\myO}_\textrm{TPC}(\vec{r}_p^+,\vec{r}_p^-)$ comprises only two points $\vec{r}_{p}^+$ and $\vec{r}_{p}^-$ of equal purity $p$, lying on different sides of the hyperplane $\myS_2$. To see this, one realizes that the (infinitesimal) purity $\delta p$ is lost while residing for a dwell time $\delta t^-$ at $\vec{r}_p^-$. To close the cycle, this purity $\delta p$ must be regained during the dwell time $\delta t^+$ at $\vec{r}_p^+$. Since the ratio of these dwell times is inverse to the ratio of the respective purity fluxes,
\begin{equation}
\frac{\delta t^+}{\delta t^-} \overset{\eqref{eq:purefluxBloch}}{=} \frac{\delta p/|f(\vec{r}_p^+)|}{\delta p/|f(\vec{r}_p^-)|}=\frac{|f(\vec{r}_p^-)|}{|f(\vec{r}_p^+)|},
\end{equation}
the time-average $\overline{\myO}$ of the TPC is given by Eq.~\eqref{eq:OTPC}. The TPC rapidly jumps back and forth between $\vec{r}_p^+$ and $\vec{r}_p^-$ via purity-preserving, unitary ``kicks'', generated by suitable, ideally $\delta$-shaped pulses $H(t)$.
In conclusion, \eqref{eq:estimation} reflects the important result that the search for the optimal asymptotic cycle can be restricted to TPCs, simplifying tremendously the original optimization over \emph{all} closed trajectories that obey \eqref{eq:purefluxBloch}.

To exemplify this reduction, we consider again the coherence $\Coh$ of a single qubit undergoing the same incoherent processes as before (with rates $\gamma_-$, $\gamma_+$, and $\gamma_d$). Due to symmetry, the most general TPC is then parametrized by two azimuthal angles and the common purity $p$, and it is constrained by $f(\vec{r}_p^+)>0$, $f(\vec{r}_p^-)<0$. In a numerical optimization of $\overline{\myO}_\textrm{TPC}(\vec{r}_p^+,\vec{r}_p^-)$, one finds that for any combination of the incoherent rates, the optimal TPC $\{\vec{r}_{*}^{+},\vec{r}_{*}^{-}\}$ degenerates to a single point, namely, the optimal stationary state $\vec{r}_{*}$ for static Hamiltonians (marked in Fig.~\ref{fig:ellipsoid}). Remarkably, this implies that no periodic Hamiltonian $H(t)$ can beat the optimal static Hamiltonian $H^*$. This is, however, a peculiarity of our choices of objective and dissipator and does not hold in general.

The same strategy can equally be applied beyond a single qubit. Similar to the static case, one obtains an upper bound for the optimum in $\mathcal{A}$ by focusing on the lowest-order set $\mathcal{A}_2$, since $\mathcal{A}\subset\mathcal{A}_2$. One can then again restrict the investigation to TPCs, rendering a numerical optimization feasible.
In our two-qubit example with spontaneous decay only, we find that the time-averaged concurrence $\overline{\Ent}=\frac{1}{T}\int_0^T \Ent(\rho(t)) \mathrm{d}t$ never exceeds the optimal static result $\Ent(\rho^*)=1/2$. This value decreases in the presence of finite absorption ($\gamma_+>0$) and/or dephasing ($\gamma_d>0$), and it never outperforms the static optimum.

\paragraph*{Conclusion.}
We developed a method to characterize the asymptotic states of open quantum systems in terms of the dynamical constraints imposed by the dissipator.
It allows us to access the optimization of arbitrary periodically time-dependent coherent control without resorting to the system Hamiltonian.
In the static case, the method leads to the characterization (4) of stabilizable states, reflecting the unitary compensability of the dissipator.
Based on that, we showed that in the general periodic case, optimizations can be restricted to the significantly simplified class of two-point cycles.  This way, optimization problems w.r.t. arbitrary objectives can be addressed that were previously prohibited by the vast range of conceivable asymptotic cycles.
To demonstrate our method, we determined the maximum asymptotic two-qubit entanglement that can be preserved by periodic coherent control in the presence of a dissipation-inducing environment.
Other relevant applications include, e.g., optimal energy transport in quantum networks \cite{Manzano:2012} or the minimization of particle loss in Bose-Einstein condensates \cite{Dalfovo:1999}.
Altogether, our method not only deepens our conceptual understanding of the working principles in open quantum systems, but also opens the prospect to treat hitherto intractable optimization problems.

\begin{acknowledgments}
S.S. acknowledges financial support by the German National Academic Foundation. A.B. acknowledges partial support through COST action MP1006. 
\end{acknowledgments}


\appendix

\section{Supplemental Material I}\label{secA:proof}

In the following, we prove that for states with nondegenerate eigenvalues criterion (\ref{eq:crit}) is sufficient for $\rho\in\myS$.
Let $\rho$ be a quantum state on a $d$-dimensional Hilbert space. If $\rho$ has nondegenerate eigenvalues, and if
\begin{equation}\label{eqA:conservespectrum}
\textrm{Tr}[\rho^{n-1} \myD(\rho)] = 0
\end{equation}
holds for $n = 2, \dots , d$, then there is a Hamiltonian $H$ that renders $\rho$ the stationary state of Eq.~(\ref{eq:Lindbladeq}), i.e.~$\rho\in\myS$.

\paragraph*{Proof:}
By assumption, $\rho=\sum_{\alpha} \lambda_\alpha \ket\alpha \bra\alpha$ has non-degenerate eigenvalues $\lambda_\alpha$. In order to prove that $\rho$ is stationary under Eq.~(\ref{eq:Lindbladeq}), we show that
\begin{equation}\label{eq:alphabetacond}
\braket{\alpha | \myD(\rho) | \beta} = \im \braket{\alpha | [H,\rho] | \beta} \quad \forall \alpha, \beta = 1,\dots, d,
\end{equation}
with the Hamiltonian $H$ as in Eq.~(\ref{eq:HgivenRho}). Since $\{\ket\alpha\}$ describes a complete basis, \eqref{eq:alphabetacond} is equivalent to the operator identity $0=\im [\rho,H]+\myD(\rho)$, and hence implies stationarity of $\rho$.

Inserting $H$ (Eq.~(\ref{eq:HgivenRho})) into \eqref{eq:alphabetacond}, and using the notation $D_{\alpha\beta} \equiv \braket{\alpha | \myD(\rho) | \beta}$, we obtain
\begin{eqnarray*}
D_{\alpha\beta} = i \sum_{\alpha'\ne\beta'}  \frac{i D_{\alpha'\beta'} } { \lambda_{\alpha'} - \lambda_{\beta'}  } \braket{\alpha |  \left( \ket{\alpha'}\bra{\beta'}\rho - \rho\ket{\alpha'}\bra{\beta'} \right)  | \beta} \\
 = \sum_{\alpha'\ne\beta'} \frac{-D_{\alpha'\beta'} } { \lambda_{\alpha'} - \lambda_{\beta'}  } (\lambda_{\beta'} - \lambda_{\alpha'})\delta_{\alpha\alpha'}\delta_{\beta'\beta} = (1-\delta_{\alpha\beta})D_{\alpha\beta}.
\end{eqnarray*}
(Note that due to the nondegeneracy of the eigenvalues $\{\lambda_\alpha\}$ we can replace $\lambda_{\alpha'} \ne \lambda_{\beta'}$ by $\alpha'\ne\beta'$ in the summation.)
The above expression is obviously true for $\alpha\ne\beta$. It remains to show that $D_{\alpha\alpha}=0$ holds for all $\alpha=1,\dots,d$. To this end, we rewrite condition \eqref{eqA:conservespectrum} as
\begin{eqnarray}\label{eqA:conservespectrum2}
0 = \textrm{Tr}[\rho^{n-1} \myD(\rho)] = \sum_{\alpha=1}^{d} (\lambda_\alpha)^{n-1} D_{\alpha\alpha}.
\end{eqnarray}
By assumption this holds for all $n=2,\dots,d$. In addition, it holds for $n=1$, since any dissipator fulfills $\textrm{Tr}[\myD(\rho)]=0$. (This can be directly verified from the Lindblad form of $\myD$). One can therefore write \eqref{eqA:conservespectrum2} as a matrix equation $\mathfrak M \,\vec{\mathfrak d} = 0 $, with $(\mathfrak M)_{\alpha n}\equiv(\lambda_\alpha)^{n-1}$, and $(\vec{\mathfrak d})_\alpha \equiv D_{\alpha\alpha}$. $\mathfrak M$ is a Vandermonde matrix, for which $\det \mathfrak M = \prod_{\alpha<\beta} (\lambda_\alpha-\lambda_\beta)$ is known \cite{Horn:1990}. Since the eigenvalues $\lambda_\alpha$ are by assumption nondegenerate, we have $\det \mathfrak M \ne 0$, and $\mathfrak M$ is hence invertible. The only solution to $\mathfrak M \,\vec{\mathfrak d} = 0$ is therefore $\vec{\mathfrak d}=0$. Thus, we have shown that $D_{\alpha\alpha}=0$ for all $\alpha=1,\dots,d$.
$\blacksquare$

\section{Supplemental Material II}
We demonstrate that no cycle can enter regions where the purity is larger than the maximal purity in $\myS_2$:
Let $p_1$ be the maximal purity in the set $\myS_2$,
\begin{equation*}
p_1 = \max_{\rho\in\myS_2} [\tr \rho^2].
\end{equation*}
Since $\myS_2$ separates the regions of positive and negative purity flux $f(\rho)$, $f(\rho)$ has the same sign for all $\rho$ with $\tr[\rho^2]>p_1$. Moreover, the purity flux must be non-positive for pure states, which are characterized by maximal purity $\tr[\rho^2]=1$; otherwise the dissipative flux would drive the states out of the space of valid quantum states. Hence, for any state with purity larger than $p_1$, the dissipative flux is negative. Let us now assume that there exists a cycle $\{\rho(t)\}$ with purity $\tr[\rho^2(t_0)]>p_1$ at some time $t_0$. The negative purity flux in this region then implies that ,going back in time, the purity must only increase. Consequently, the state trajectory will never leave the region of negative flux, which contradicts our assumption that $\{\rho(t)\}$ is a cycle. This proves that states with purity larger than $p_1$ can never be traversed in an asymptotic cycle.

Corresponding statements hold also for the higher moments $\tr[\rho^n]$.

\end{document}